%% file: ANA-HION-2025-16-PAPER.tex
\documentclass[PAPER, atlasdraft=false, cernpreprint, USenglish, texmf, orcidlogo]{atlasdoc}
\usepackage{atlaspackage}

\usepackage{atlasbiblatex}

\usepackage{atlasphysics}
\graphicspath{{logos/}{figures/}}

\AtlasTitle{Observation of centrality-dependent dijet transverse momentum imbalance in O+O and Ne+Ne collisions at \ensuremath{\sqrt{s_{_\text{NN}}}} = 5.36 TeV with the ATLAS detector}
\AtlasAbstract{%
The ATLAS experiment presents an observation of a centrality-dependent dijet transverse momentum imbalance in O+O and Ne+Ne collisions at a nucleon–nucleon center-of-mass energy of 5.36 TeV at the Large Hadron Collider.
The measurement uses 8.0 nb$^{-1}$ of O+O and 1.0 nb$^{-1}$ of Ne+Ne data collected in 2025, together with 386 pb$^{-1}$ of \textit{pp} data at the same energy used as a reference.
The dijet momentum balance is quantified using the ratio of the sub-leading jet transverse momentum to that of the leading jet, $x_J$.
For dijets produced azimuthally back-to-back, the self-normalized $x_J$ distributions exhibit increasingly large deviations from the \textit{pp} reference as collisions become more central, corresponding to an increasing overlap of the colliding nuclei.
The observed centrality dependence is consistent with medium-induced partonic energy loss in O+O and Ne+Ne collisions, demonstrating that such effects persist in collision systems considerably smaller than Pb+Pb and Xe+Xe. These results establish a new regime for investigating the path-length dependence of jet quenching and constrain the onset of quark--gluon plasma effects in small nuclear collision systems. }
\PreprintIdNumber{CERN-EP-2026-178}

\AtlasJournal{Phys.\ Rev.\ Lett.}
\hypersetup{pdftitle={ATLAS document},pdfauthor={The ATLAS Collaboration}}

\clearpage
\newcommand{\CSOO}{{10.2$\sigma$}}
\newcommand{\CSNeNe}{{8.9$\sigma$}}
\newcommand{\MMCOO}{{7.4$\sigma$}}
\newcommand{\MMCNeNe}{{5.3$\sigma$}}
\newcommand{\MCSOO}{{5.8$\sigma$}}
\newcommand{\MCSNeNe}{{4.9$\sigma$}}

\newcommand{\RFour}{\mbox{$R = 0.4$}}

\newcommand{\PT}{$p_{\text{T}}$}

\newcommand{\PTONE}{\mbox{\ensuremath{p_{\text{T}_\text{1}}}}}
\newcommand{\PTTWO}{\mbox{\ensuremath{p_{\text{T}_\text{2}}}}}

\newcommand{\PTTONE}{p_{\text{T}_\text{1}}^{\text{truth}}}
\newcommand{\PTTTWO}{p_{\text{T}_\text{2}}^{\text{truth}}}

\newcommand{\TAB}{\ensuremath{T_{\text{AB}}}}

\newcommand{\TAA}{$\langle T_{\text{AA}} \rangle$}

\newcommand{\NCOLL}{$\langle N_{\text{coll}} \rangle$}

\newcommand{\diff}{\ensuremath{\text{d}}}
\newcommand{\Njet}{\ensuremath{N_{\mathrm{jet}}}}

\newcommand{\pbpb}{Pb$+$Pb}

\newcommand{\OO}{O$+$O}
\newcommand{\NeNe}{Ne$+$Ne}

\newcommand{\XJ}{\mbox{$x_{\text{J}}$}}

\newcommand{\SNN}{\ensuremath{\sqrt{s_{_\text{NN}}}}}

\newcommand{\pythia}{\mbox{\textsc{Pythia}8}}

\newcommand{\pxp}{$pp$}

\newcommand{\OxO}{O+O}

\newcommand{\fcalsum}{\ensuremath{\Sigma E_\mathrm{T}^\mathrm{FCal}}\xspace}
\clearpage

\begin{document}

\maketitle

%

%
%

%
%
%
%

%
\newcommand{\AtlasCoordFootnote}{%
ATLAS uses a right-handed coordinate system with its origin at the nominal interaction point (IP)
in the center of the detector and the \(z\)-axis along the beam pipe.
The \(x\)-axis points from the IP to the center of the LHC ring,
and the \(y\)-axis points upwards.
Polar coordinates \((r,\phi)\) are used in the transverse plane,
\(\phi\) being the azimuthal angle around the \(z\)-axis.
The pseudorapidity is defined in terms of the polar angle \(\theta\) as \(\eta = -\ln \tan(\theta/2)\) and is equal to the rapidity
$ y = \frac{1}{2} \ln \left( \frac{E + p_z}{E - p_z} \right) $ in the relativistic limit.
Angular distance is measured in units of \(\Delta R \equiv \sqrt{(\Delta y)^{2} + (\Delta\phi)^{2}}\).}

One of the primary goals of the heavy-ion program at the Large Hadron Collider (LHC) %
is to characterize the quark--gluon plasma (QGP), a deconfined state of strongly interacting matter produced at extreme temperature and density~\cite{Busza:2018rrf}. High-transverse-momentum (\PT) jets provide a powerful probe of the QGP over a broad range of resolution scales~\cite{Cunqueiro:2021wls}. As energetic partons traverse the medium, they undergo \textit{jet quenching}, manifested through a reduction of the jet energy and a redistribution of energy to large angles via collisional and medium-induced radiative processes~\cite{review:jet_quenching}. The magnitude of the energy loss is expected to depend on the medium properties, including density and temperature, and the in-medium path length traversed by the parton~\cite{Gyulassy:1993hr, Baier:1996kr, Zakharov:1997uu, Wiedemann:2000za, Gyulassy:2000gk, Betz_2014, Arleo:2022shs}, with additional jet-by-jet fluctuations arising from the stochastic nature of the energy-loss process~\cite{Milhano:2015mng}.

Evidence for path-length-dependent energy loss has been obtained in non-central Pb+Pb collisions through azimuthal modulations of high-\pt\ particle and jet production relative to the event plane~\cite{HION-2012-12,ALICE:2015efi,HION-2020-09}. However, fluctuations in the initial-state nucleon configuration and energy deposition complicate the determination of the effective path length, particularly in peripheral collisions. In proton--nucleus collisions, signatures of collective behavior have been observed~\cite{Nagle:2018nvi}, while unambiguous evidence for jet quenching remains elusive~\cite{HION-2021-17,CMS-HIN-23-010}.

Collisions of light ions provide a new opportunity to investigate the path-length dependence of jet quenching in systems with a substantially smaller transverse extent than \pbpb\ collisions. Central \OO\ and \NeNe\ collisions retain an extended and approximately symmetric transverse overlap geometry while reducing the characteristic in-medium path length relative to central \pbpb\ collisions, for which the typical path length is expected to be approximately a factor of two larger. This results in a correspondingly smaller expected magnitude of energy loss~\cite{Huss:2020whe}. Furthermore, the more symmetric overlap geometry of central light-ion collisions relative to peripheral \pbpb\ interactions is expected to reduce geometric uncertainties associated with the interpretation of path-length-dependent effects~\cite{Loizides:2025ule}.

Recent measurements~\cite{HION-2025-02,ALICE:2025luc,CMS-HIN-25-009}
show evidence of collective behavior in
both \OO\ and \NeNe\ collisions, suggesting the formation of a QGP.
Complementary measurements of hard probes have also been used to search for signatures of medium-induced effects in light-ion collisions.
The CMS Collaboration measured the charged-hadron nuclear modification factor in centrality-integrated \OO~\cite{CMS-HIN-25-008} and \NeNe~\cite{CMS-HIN-25-014} collisions at $\sqrt{s_{\mathrm{NN}}}=5.36$~\TeV\ and observed a suppression relative to unity that is not described by calculations including only nuclear modifications to parton distributions. Additionally, the STAR Collaboration at RHIC observed a suppression of hadron--hadron and hadron--jet correlations in central \OO\ collisions relative to a peripheral reference at $\sqrt{s_{\mathrm{NN}}}=200$~\GeV~\cite{STAR:IAA_OO}. Both of these measurements are suggestive of partonic energy loss in light-ion collisions.

In small collision systems, the interpretation of jet-quenching observables is complicated by correlations between hard-process production and the soft underlying-event activity used to characterize the collision geometry~\cite{Loizides:2017sqq,PHENIX:2013jxf,Park:2025mbt}. These ``centrality bias'' effects can become sizable in measurements based on absolute jet yields or Glauber-model scaling quantities such as \TAA\ or \NCOLL~\cite{ALICE:2014xsp,HION-2012-14,HION-2023-05, Alvioli:2017wou}. Measurements that do not depend on absolute rates or quantities extracted from Glauber modeling and are based on distributions normalized to unit area are expected to be significantly less sensitive to such effects.

In Quantum Chromodynamics (QCD), jet production proceeds through the fragmentation of quarks and gluons from a short-distance scattering process.
At leading order, pairs of jets (dijets) are produced approximately balanced in the plane transverse to the beam (back-to-back).
Dijet observables are sensitive to path-length-dependent energy loss, since the two jets may traverse different distances through the medium~\cite{Qin:2010mn}.
Additionally, they are less sensitive than jet yields to centrality-bias effects~\cite{Park:2025mbt} and to uncertainties associated with nuclear parton distributions~\cite{Gebhard:2024flv}, which remain poorly constrained for oxygen and neon nuclei~\cite{PhysRevD.105.L031504}. These features make dijets a particularly powerful probe of medium-induced modifications in light-ion collisions.

The quantity used to characterize the quenching of the two jets in the dijet system is the dijet momentum balance,
\begin{equation}
\XJ = \frac{\PTTWO}{\PTONE},
\end{equation}
where $\PTONE$ and $\PTTWO$ are the transverse momenta of the leading and sub-leading jets, respectively, that are required to be nearly back-to-back in azimuth with $\Delta\phi > 7\pi/8$.
Measurements of $\XJ$ at the LHC, beginning with the first observations of enhanced dijet asymmetry in central \pbpb\ collisions at $\sqrt{s_{\mathrm{NN}}}=2.76$~\TeV~\cite{HION-2010-02,CMS-HIN-10-004}, established the dijet momentum balance as a sensitive probe of partonic energy loss. Subsequent ATLAS measurements~\cite{HION-2012-11,HION-2019-02,HION-2018-28} at 2.76 and 5.02~\TeV\ incorporated unfolding techniques to correct for detector effects and jet energy resolution, as well as scanning systematically the results as a function of the jet radius~\cite{HION-2024-03}.

Measurements of dijet \XJ\ distributions in \OO\ and \NeNe\ collisions recorded by ATLAS in 2025 are presented as a function of centrality and compared with \pp\ collisions at the same center-of-mass energy.
The results are presented in the form of ``self-normalized'' jet-pair distributions corrected for detector effects,
\begin{equation}
\frac{1}{N_\mathrm{pair}}\frac{\diff N_\mathrm{pair}}{\diff \XJ},
\label{eq:self_normalized_xj}
\end{equation}
where $N_\mathrm{pair}$ is the number of jet pairs. Centrality is defined through total transverse energy measured in the ATLAS forward calorimeters.
Results are reported for $63 < \PTONE < 251$~\GeV\ and $0.32 < \XJ < 1$ using jets reconstructed within $|y| < 2.1$.\footnote{\AtlasCoordFootnote} The \XJ\ and rapidity selections follow previous ATLAS dijet measurements~\cite{HION-2012-11,HION-2019-02,HION-2018-28,HION-2024-03}. Owing to the significantly reduced underlying-event contribution in light-ion collisions, the measurement extends to lower jet transverse momentum than in previous heavy-ion measurements. Comparisons with different theoretical predictions are also presented. %

%
%

%
%
%
%


%

%
%
%
%

%
%
%
%
The ATLAS detector~\cite{GENR-2019-02, PERF-2007-01} is a multipurpose particle physics detector with cylindrical geometry.
The measurement presented here is performed using the ATLAS tracking detector, calorimeters, trigger, and data acquisition systems.
The inner tracking
detector covers the pseudorapidity range \(|\eta| < 2.5\) and consists of silicon pixel, silicon microstrip, and
transition radiation tracking detectors all immersed in a \qty{2}{\tesla} axial magnetic field.
The calorimeter system consists of a sampling liquid-argon (LAr) electromagnetic (EM) calorimeter covering $|\eta| < 3.2$, a steel--scintillator sampling hadronic calorimeter covering $|\eta| < 1.7$, LAr hadronic calorimeters covering $1.5 < |\eta| < 3.2$,
and a LAr~Forward Calorimeter (FCal) covering $3.2 < |\eta| < 4.9$.
Forward neutrons produced from
the breakup of nuclei in both hadronic and electromagnetic interactions
are measured by compact tungsten sampling Zero Degree Calorimeters (ZDCs) \cite{Avoni:2025lmu} positioned
at $z=\pm140$ meters from the ATLAS interaction point.
The luminosity is measured mainly by the LUCID--2 detector \cite{LUCID2} that records Cherenkov light produced in the quartz windows of photomultipliers located close to the beampipe.

Events were selected by the first-level trigger system implemented in custom hardware,
followed by selections made by algorithms implemented in software in the high-level trigger (HLT)~\cite{TRIG-2022-01}.
A software suite~\cite{SOFT-2022-02} is used in data simulation, in the reconstruction
and analysis of real and simulated data, in detector operations, and in the trigger and data acquisition systems of the experiment.


The analysis presented in this Letter uses 8~\inb\ of \OO\ and 1~\inb\ of \NeNe\ collision data collected at \SNN = 5.36~\TeV\ with the ATLAS detector in 2025, together with 386~\ipb\ of \pp\ collision data recorded in 2024 at the same center-of-mass energy~\cite{DAPR-2021-01}.
Events used in the analysis are required to be recorded during stable running conditions of the LHC and to have no detector hardware or readout errors.
Events are selected using a combination of minimum-bias (MB) and jet triggers~\cite{TRIG-2022-01}.
The MB trigger requires at least one reconstructed track with $\pt > 200$~\MeV~\cite{ATL-INDET-PUB-2009-002}.
Jet triggers in Run~3 use the upgraded Level-1 jet Feature EXtractor (jFEX) system~\cite{Bauss:2018nde}, while the HLT employs the same jet reconstruction algorithm used offline.
In both light-ion collision systems, two single-jet triggers with transverse-energy (\ET) thresholds of 20 and 40~\GeV\ at the HLT are used.
In \pp\ collisions, several jet triggers with thresholds between 20 and 120~\GeV\ are employed.
The highest-threshold jet triggers samples the full \pp, \OO\, and \NeNe\ luminosity. All leading jets included in the analysis lie within the kinematic region where the corresponding jet triggers are more than 99\% efficient.

As in previous ATLAS heavy-ion measurements~\cite{HION-2011-01,HION-2011-05,HION-2019-03}, events are categorized into centrality intervals using the total transverse energy measured in the FCal, \fcalsum.
The relation between the \fcalsum\ distribution and the sampled fraction of the total inelastic \OO\ and \NeNe\ cross-sections is determined using a Monte Carlo Glauber model calculation~\cite{Miller:2007ri,Loizides:2025ule}.
Five centrality intervals are defined: 0--10\%, 10--20\%, 20--40\%, 40--60\%, and 60--80\%, where smaller percentiles correspond to more central collisions. More details on the centrality determination methodology used can be found in Ref.~\cite{ATLAS-CONF-2026-004}.
Events consistent with photo-nuclear or diffractive topologies are suppressed by requiring at least one topo-cluster~\cite{PERF-2014-07} with transverse energy greater than 0.4~\GeV\ on each side of the forward calorimeter.

The recorded \OO\ and \NeNe\ data contain non-negligible contributions from pileup events, in which two or more inelastic collisions occur in the same bunch crossing. Because the event activity from multiple collisions is superimposed, the centrality of such events cannot be determined reliably. These events are therefore removed from the analysis.
Most pileup events are removed by requiring each event to contain only one high-quality reconstructed vertex~\cite{IDTR-2021-01}, defined as one that is well-constrained in position and characterized by many associated tracks pointing to a common origin; such vertices are selected by requiring a $z$-position variance of less than $0.02~\mathrm{mm}^2$.
Correlations between the ZDC energy and \fcalsum\
are used to suppress additional pileup by rejecting events which
deviate significantly from the expected correlations, assuming a Gaussian distribution for each interval in \fcalsum.
Additional beam-induced background is suppressed by removing events with track multiplicities significantly smaller than expected relative to \fcalsum.

Collisions affected by pileup are not rejected in \pp\ data, where jets originating from pileup vertices are rejected by requiring that tracks matched to the jet are consistent with originating from the primary vertex~\cite{PERF-2014-03}.

The same jet reconstruction procedure is used for the \OO, \NeNe, and \pp\ collision data, following that used in previous ATLAS measurements in \pbpb\ collisions~\cite{HION-2017-10,HION-2020-09}, with settings re-optimized for the lower-activity environment of light-ion collisions. Jets are reconstructed from calorimeter towers of size $\Delta\eta \times \Delta\phi = 0.1 \times \pi/32$ using the \antikt\ algorithm~\cite{Cacciari:2008gp} implemented in FastJet~\cite{Cacciari:2011ma} and a radius parameter of \RFour.

An event-by-event background subtraction procedure uses estimates of the average underlying-event transverse energy density, $\rho(\eta)$, to evaluate the background contribution to the jet kinematics.
The impact of jet energies on the estimated $\rho$ values is mitigated by excluding jets through an iterative procedure~\cite{ATLAS-CONF-2015-016}.
In contrast to \pbpb\ collisions, the azimuthal modulation of the underlying-event background arising from collective particle flow is neglected, since its effect on jet reconstruction is negligible for the jet kinematics considered.

Simulation-based $\eta$- and \pT-dependent corrections are applied to measured jets to calibrate the jet energy response and jet energy resolution~\cite{ATLAS:2014hvo,ATLAS-CONF-2015-016}. Additional corrections derived from \textit{in situ} studies of $\gamma$+jet, $Z$+jet, and dijet events are applied to account for residual differences between data and simulation~\cite{HION-2018-06}. The absolute jet energy scale (JES), the $\eta$-intercalibration, and the jet energy resolution (JER) are derived from in situ studies using Run~3 \pxp\ data at $\sqrt{s}=13.6$~\TeV. A cross-calibration then transfers the absolute JES from high-luminosity 13.6~\TeV\ collisions to low-luminosity 5.36~\TeV\ data~\cite{ATLAS-CONF-2015-016}.

To evaluate the jet reconstruction performance and detector response, dijet Monte Carlo samples were simulated using \pythia~\cite{Sjostrand:2014zea} with the A14 tune~\cite{ATL-PHYS-PUB-2014-021} and the NNPDF23LO parton distribution function set~\cite{Ball:2012cx}.
The detector response was modeled using the full ATLAS detector simulation based on the \textsc{Geant4} toolkit~\cite{AGOSTINELLI2003250}.
For \OO\ and \NeNe\ collisions, the \textsc{Geant4} simulation output from \pythia\ events were overlaid with minimum-bias data events of the same collision system to reproduce the underlying-event conditions observed in data.
These ``data overlay'' samples were recorded using a combination of \ET-based triggers to enhance the fraction of central collisions and, at the level of analysis, were reweighted on an event-by-event basis to match the centrality distribution of the jet-triggered data sample.


%
%
%


The analysis and dijet selections closely follow the previous ATLAS measurement in \pbpb\ collisions~\cite{HION-2019-02}, to facilitate comparisons between the results.
The leading dijet pair is formed from the two highest-\pt\ jets in the event. This jet pair must satisfy the fiducial dijet selections that require reconstructed $\PTONE > 50$~\GeV, $\PTTWO > 20$~\GeV, $|y|<2.1$, and $\Delta\phi > 7\pi/8$.
Results are reported for $63 < \PTONE < 251$~\GeV\ and $\XJ > 0.32$, while events outside this range are retained in underflow bins for the unfolding procedure.
For all collision systems, \XJ\ is measured in the same range, with five bins in \NeNe\ and ten bins in \OO.
The binning in \XJ\ was chosen to be coarser in \NeNe\ collisions to account for the smaller data sample size.
For comparisons to \OO\ (\NeNe), the \pp\ reference is measured using the same ten-bin (five-bin) \XJ\ binning.

The measured $(\PTONE,\PTTWO)$ distributions include contributions from both correlated and uncorrelated jet pairs. The latter arise from independent hard scatterings involving different nucleon pairs within the same ion collision and are uniformly distributed in $\Delta\phi$. Their contribution is estimated by using a sideband method.
A symmetrized $(\PTONE,\PTTWO)$ distribution constructed from dijets satisfying $0.8 < \Delta\phi < 1.1$ is normalized to the signal region and subtracted from the data.
The presence of uncorrelated jets also introduces an inefficiency for genuine dijets when one jet in the pair is replaced by an uncorrelated third jet.
This effect is corrected using the measured inclusive jet spectrum in minimum-bias events following the procedure of Ref.~\cite{HION-2012-11}.
The combination of the sideband subtraction and the correction for the uncorrelated jet inefficiency results in a net correction of at most 3.2\% to the final \XJ\ distributions in the most central 0--10\% interval, and significantly less in more peripheral collisions.
Contributions from three-jet topologies to the sideband region are found to be less than 0.5\% in simulation and neglected.
Potential contributions from the azimuthal anisotropy of particle production ("harmonic flow"), which can induce a modulation of the uncorrelated dijet background as a function of $\Delta\phi$, are evaluated through the centrality dependence of the $\Delta\phi$ sideband modulation~\cite{HION-2012-11}.
The resulting effect on the measurement is found to be negligible in \OO\ and \NeNe\ collisions.

To correct for detector effects, including any residual JES deviation from unity, the finite JER, and reconstruction inefficiencies, the measured $(\PTONE,\PTTWO)$ distributions are unfolded using an iterative Bayesian procedure~\cite{DAgostini:1994fjx} implemented in RooUnfold~\cite{Adye:2011gm}. A two-dimensional unfolding is performed to account for bin migration of both jets and possible interchange of the leading and sub-leading jet ordering. The response matrices include migrations into and out of the fiducial region and are symmetrized in $(\PTTONE,\PTTTWO)$ to regularize the unfolding procedure. Separate response matrices are derived for \pp\ and each light-ion centrality interval using simulated dijet events with truth and reconstructed jets matched within $\Delta R<0.3$. Truth jets are defined as in Ref.~\cite{HION-2019-02}.

The unfolding prior is taken from the \pythia\ simulation. The number of iterations is optimized independently for \pp, \OO, and \NeNe\ collisions to balance residual bias against statistical fluctuations, resulting in four iterations for the \pp\ measurement and for the \OO\ and \NeNe\ 0--10\% centrality intervals, and in three iterations for all other centralities. Statistical uncertainties are evaluated using bootstrap variations of both the data and response matrices~\cite{ATL-PHYS-PUB-2021-011}.
The unfolded $(\PTONE,\PTTWO)$ distributions are reflected about $\PTONE=\PTTWO$ to restore the leading/sub-leading jet hierarchy and projected into bins of the dijet momentum balance observable, \XJ, following the procedures used in Refs.~\cite{HION-2012-11,HION-2019-02,HION-2018-28}.

In addition to biases arising from correlations between hard-scattering and soft particle production, the centrality determination could also be biased directly by particles produced in the hard process itself including, for example, the presence of additional jets within the acceptance of the centrality estimator.  %
This effect is studied using simulated \pythia\ dijet events overlaid with events generated with \textsc{Hijing}~\cite{Wang:1991hta} and processed with the same centrality selection as used in data.
Since the dijet signal is generated with unquenched \pythia\ in both \pp\ and light-ion collisions, any modification to the \XJ\ distributions relative to \pp\ would indicate a topology-driven bias coming from the centrality determination.
Deviations relative to \pp\ are found to be small and generally comparable to the statistical uncertainties, indicating that multijet final states do not significantly affect the measured \XJ\ distributions or the interpretation of the results.


%

%
%
%

Systematic uncertainties in the measured \XJ\ distributions arise from the JES, JER, combinatorial background subtraction, and unfolding procedure.
The JES uncertainty includes contributions from \textit{in situ} calibration, cross-calibration, flavor response, flavor composition, and, for \OO\ and \NeNe\ collisions, possible modifications of the calorimetric response due to medium-modified parton showers~\cite{HION-2019-02,ATLAS-CONF-2015-016}. The JER uncertainty is evaluated by varying the jet resolution in simulation according to \textit{in situ} measurements based on dijet momentum balance in \pp\ collisions~\cite{ATLAS:2020cli}. The use of Monte Carlo samples with data overlay avoids the need to assign an additional centrality-dependent uncertainty to account for residual differences between underlying-event fluctuations in data and simulation. The resulting JER variations are symmetrized. Their impact is evaluated by repeating the full analysis with modified response matrices and comparing the resulting self-normalized \XJ\ distributions with the nominal result. An uncertainty in the combinatorial background subtraction is evaluated by varying the $\Delta \phi$ window that defines the sideband region.  The nominal window is varied  ($1.1<\Delta\phi < 1.4$) and the full analysis is repeated.  The resulting difference is symmetrized and taken as a systematic uncertainty.

Additional systematic uncertainties arise from the unfolding procedure. The sensitivity to the Bayesian prior is evaluated by reweighting the prior to approximately reproduce the reconstructed distributions in data and repeating the unfolding. The difference from the nominal unfolded result is symmetrized and assigned as a systematic uncertainty. An additional systematic uncertainty is assigned to account for the closure after experimental effects were re-applied to the unfolded result by means of the response matrix. The closure is consistent with unity within statistical uncertainties, which are mostly below 3\% in \pp\ collisions and increase in \OO\ and \NeNe\ collisions, particularly at low \XJ\ where they can reach 15--20\%. Any deviations from unity are conservatively taken as an additional systematic uncertainty.

Generally, the JES and the JER dominate the total systematic uncertainties. For most bins of \XJ\ the total systematic uncertainty is below 10--15\%, only reaching higher values in the lowest bins of \XJ. At low \PTONE\ the measurement is dominated by systematic uncertainties, while at high \PTONE\ the statistical uncertainties become more significant.
%


%

%
%
%

\begin{figure}[ht]
\centering

\begin{minipage}{0.48\linewidth}
\centering
\includegraphics[width=\linewidth,trim={0cm 0cm 0cm 0mm},clip]{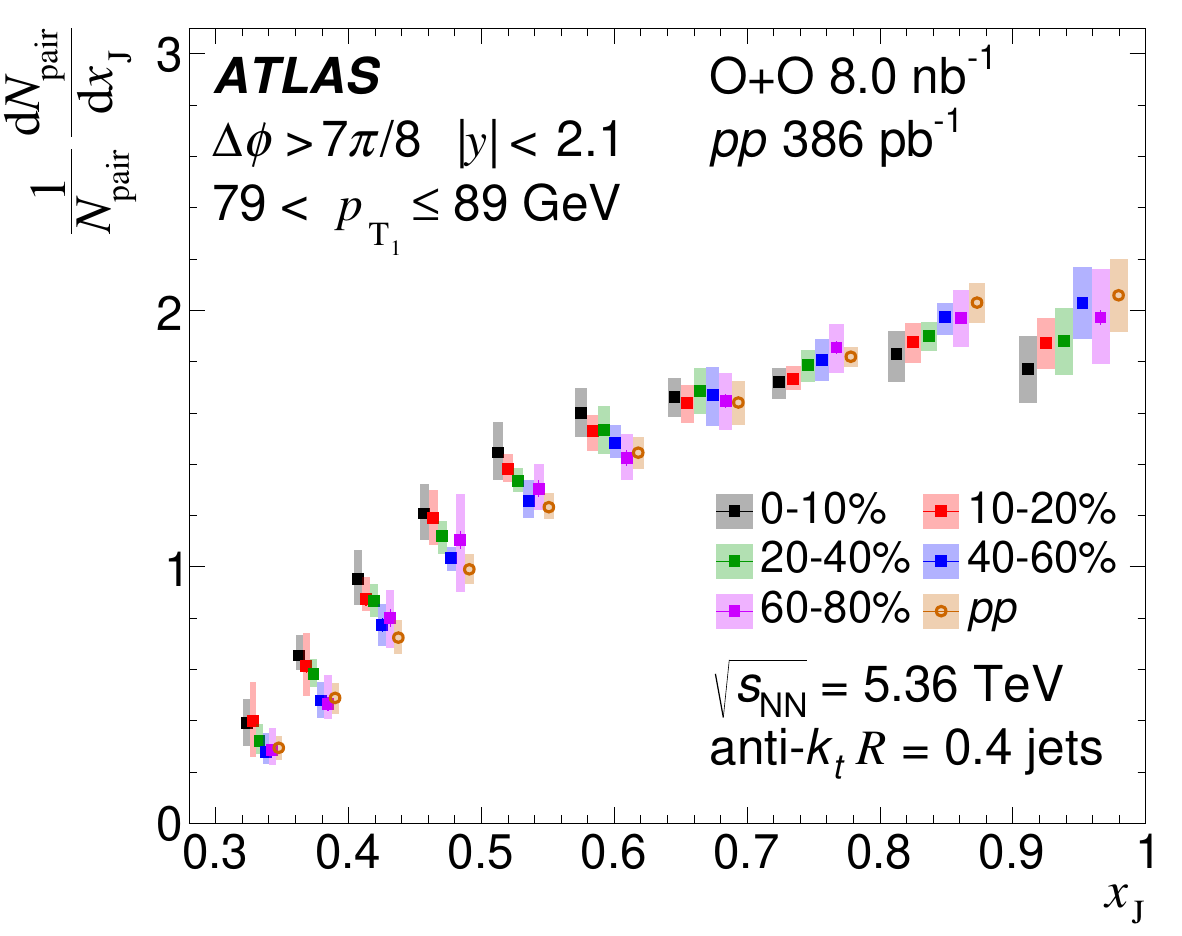}
(a)\\[-0.5ex]
\end{minipage}
\hfill
\begin{minipage}{0.48\linewidth}
\centering
\includegraphics[width=\linewidth,trim={0cm 0cm 0cm 0mm},clip]{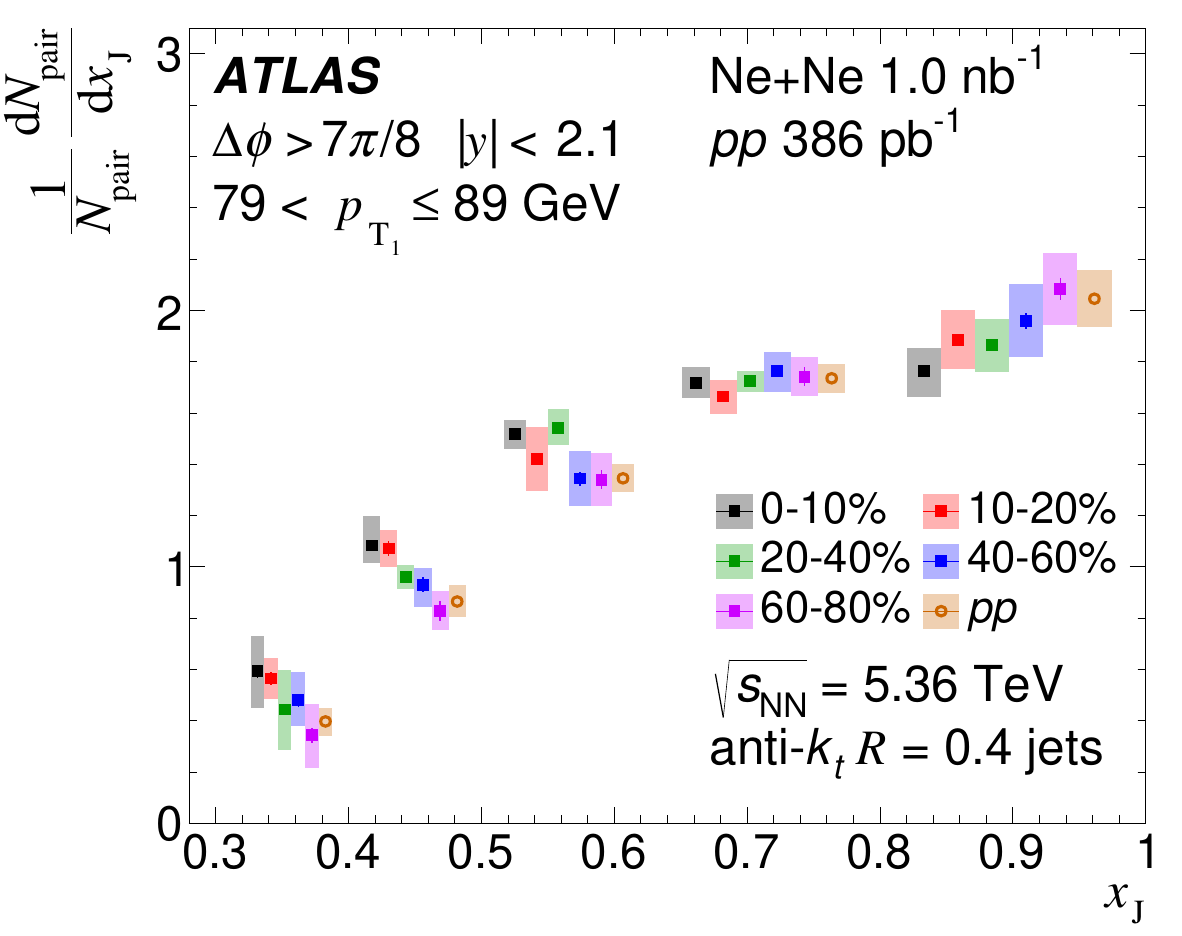}
(b)\\[-0.5ex]
\end{minipage}

\caption{
Unfolded self-normalized \XJ\ distributions in
(a) \OxO\ and (b) \NeNe\ collisions, compared with the
\pxp\ reference, in a selected bin of leading-jet \PT\
for all centrality selections. Points in the same \XJ\ selection
are horizontally offset for visual clarity. Vertical error bars represent the statistical uncertainty, and the shaded areas represent the total uncertainty.
}
\label{fig:xj_fullcent_result}
\end{figure}

Figure~\ref{fig:xj_fullcent_result} shows the measured \XJ\ distributions, given by Eqn.~\ref{eq:self_normalized_xj}, for all centrality bins in \OO\ and \NeNe\ collisions for $79 < \PTONE \leq 89$~\GeV. In both light-ion collision systems, the modification of the \XJ\ distributions relative to \pxp\ is observed to have systematic centrality dependence, with a clear trend of increasing modification with increasing event activity.
The trend observed in Figure~\ref{fig:xj_fullcent_result} is qualitatively compatible with the formation of QGP droplets of increasing transverse size inducing greater energy-loss effects on colored probes formed in the hard-scattering.  In the 0--10\% central \OO\ and \NeNe\ collisions, the \XJ\ modification is found to be similar to that observed in \pbpb\
collisions with a similar level of event activity. The comparison is reported in Appendix A in the End Matter.

To quantify the modification of the \XJ\ distributions in \OO\ and \NeNe\ collisions, their ratios to those measured in the \pp\ reference are evaluated for each centrality interval. Uncertainties in the ratio arising from the JES, JER, and combinatorial background subtraction are correlated, and thus partially cancel out in the ratio.  The significance of the deviation of the ratios from a null hypothesis of no modification (unity) is determined using a $\chi^2$ test based on the full covariance matrix of the self-normalized ratios, including bin-to-bin correlations from statistical and systematic uncertainties. %
For asymmetric uncertainties, the covariance matrix corresponding to the uncertainty variation that yields the smallest deviation from unity is used, resulting in a conservative estimate of the significance.

\begin{figure}[ht]
\centering
\includegraphics[width=0.98\linewidth,trim={0cm 0cm 0cm 0mm},clip]{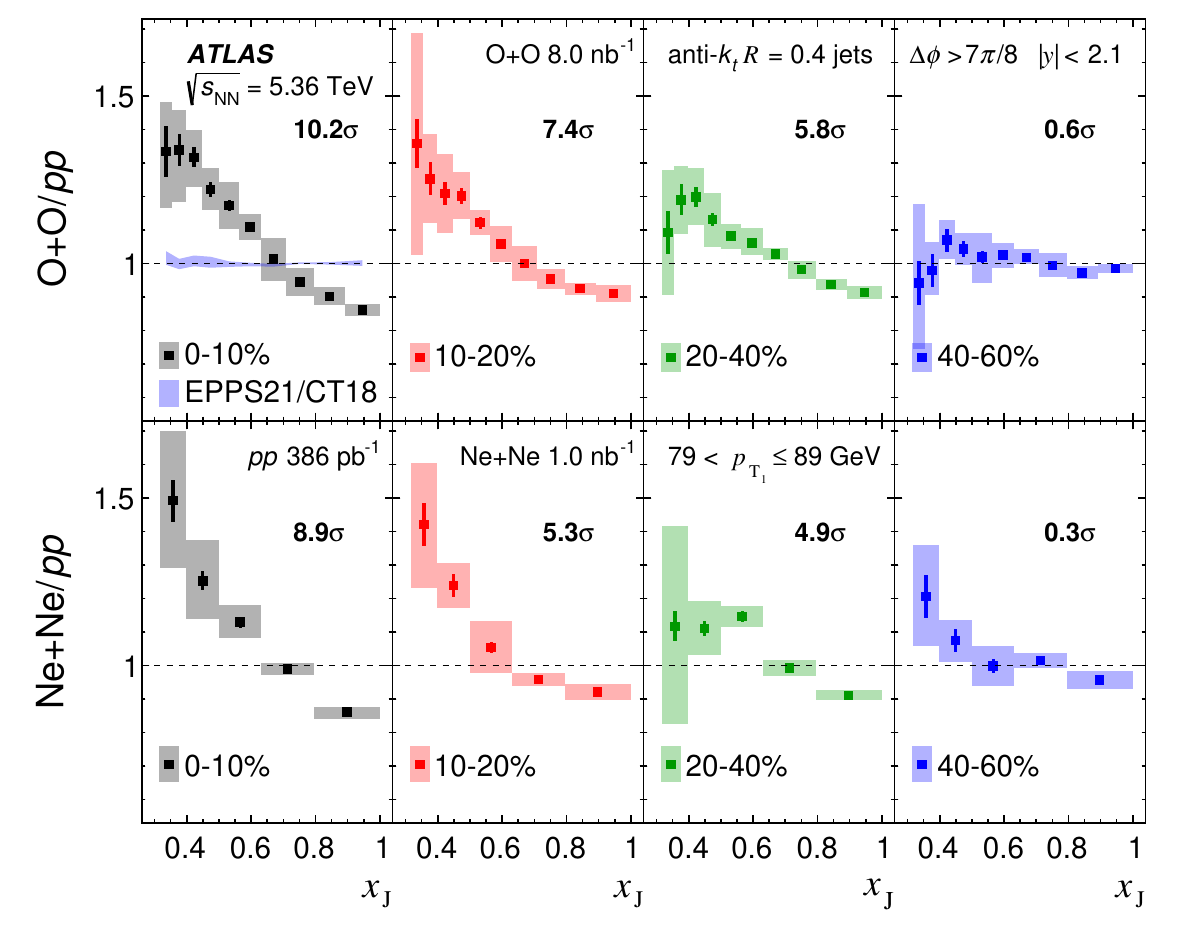}

\caption{
Ratios of unfolded self-normalized \XJ\ distributions in \OxO~(top) and \NeNe~(bottom) collisions divided by those in \pxp\ collisions for
79~$< \PTONE \leq 89$~\GeV\ in the centrality intervals 0--10\% (left), 10--20\% (left-middle),  20--40\% (right-middle), and 40--60\% (right). Vertical error bars represent the statistical uncertainty, and the shaded areas represent the total uncertainty.
The 0--10\% \OxO\ panel also shows theoretical predictions for the centrality-integrated non-quenching baseline, including only nPDF effects. These are evaluated from the ratio of truth-level \XJ\ distributions obtained with the EPPS21~\cite{Eskola:2021nhw} nPDF set and the CT18~\cite{Hou:2019efy} proton PDF set. The shaded band represents the theoretical uncertainty in the prediction.
}
\label{fig:xj_ratio_significant}
\end{figure}

Figure~\ref{fig:xj_ratio_significant} presents the ratios to \pp\ for the 0--10\%, 10--20\%, 20--40\%, and 40--60\% centrality intervals.
In the interval of leading jet \PT\ between 79 and 89~\GeV\ the significance of the modification is found to be \CSOO\ (\CSNeNe) in 0--10\% central collisions, \MMCOO\ (\MMCNeNe) in 10--20\% central collisions, and \MCSOO\ (\MCSNeNe) in 20--40\% mid-central collisions for \OO\ (\NeNe).
In the same leading jet \PT\ bin, the modification is not significant in more peripheral collisions (40--60\% and 60--80\%) in either system. A significant modification (>5$\sigma$) is observed in \OO\ (\NeNe) for collisions in the 0--10\% centrality range for \PTONE\ < 141 (126)~\GeV.
More details on the methodology applied for the significance analysis and tables with the significance assessed in each analysis bin are reported in Appendix B in the End Matter.
Figure~\ref{fig:xj_ratio_significant} also shows that the modifications expected solely from nuclear PDF (nPDF) and isospin effects, evaluated using the full EPPS21~\cite{Eskola:2021nhw} oxygen nPDF set and the CT18~\cite{Hou:2019efy} proton PDF set, are incompatible with the observed modification. Although the comparison is shown only for the 0--10\% centrality interval, the predicted nPDF effects are obtained from the full nuclear PDF modification and are not specific to that centrality selection. Across the full \PTONE\ range, the nPDF effects remain within 4\% of unity and do not show a systematic dependence on either \PTONE\ or \XJ. A cross-check using nCTEQ15~\cite{Kovarik:2015cma} instead of EPPS21 yielded similar results.

\begin{figure}[ht]
\centering

\begin{minipage}{0.43\linewidth}
\centering
\includegraphics[width=\linewidth,
trim={0cm 0cm 0cm 0mm},clip]{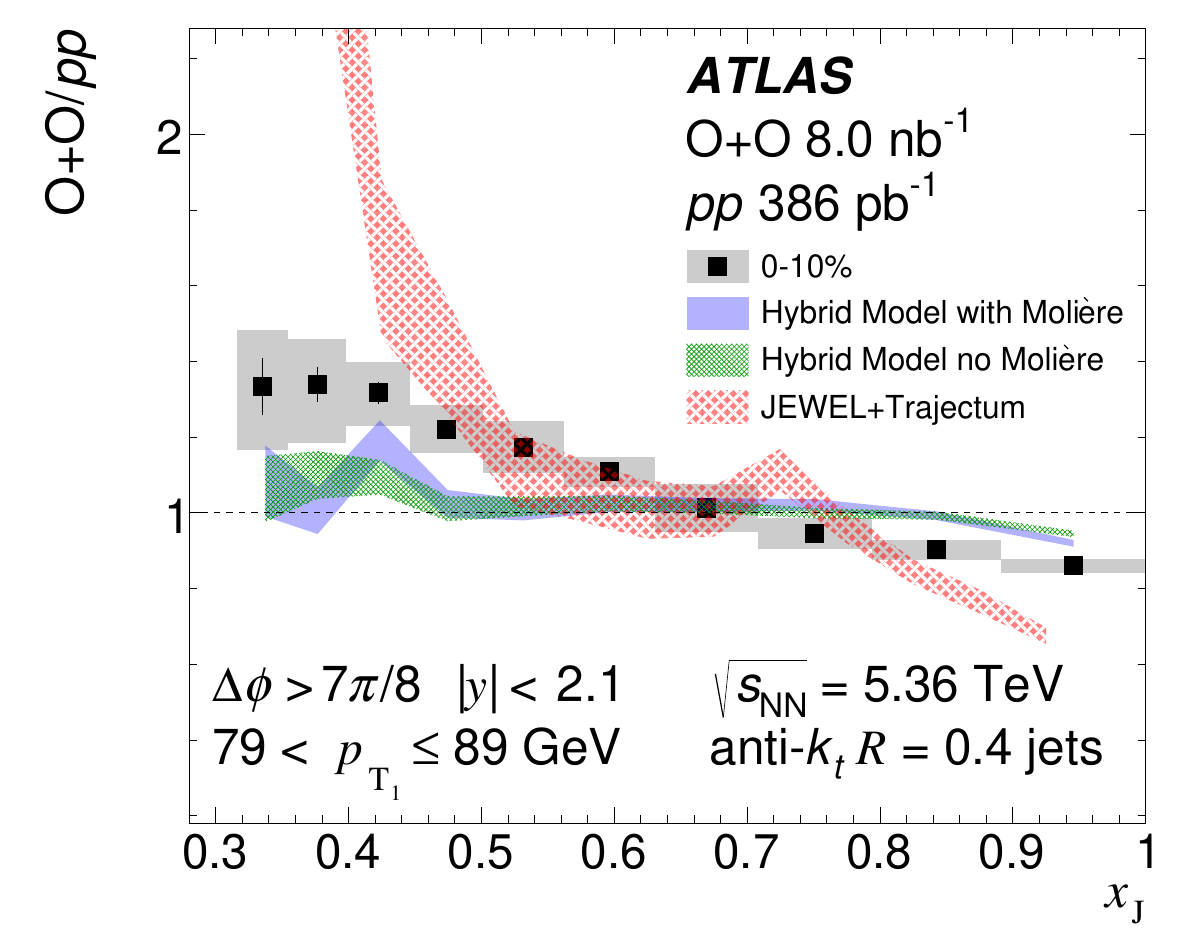}
(a)\\[-1ex]
\end{minipage}
\hspace{0.4cm}
\begin{minipage}{0.43\linewidth}
\centering
\includegraphics[width=\linewidth,
trim={0cm 0cm 0cm 0mm},clip]{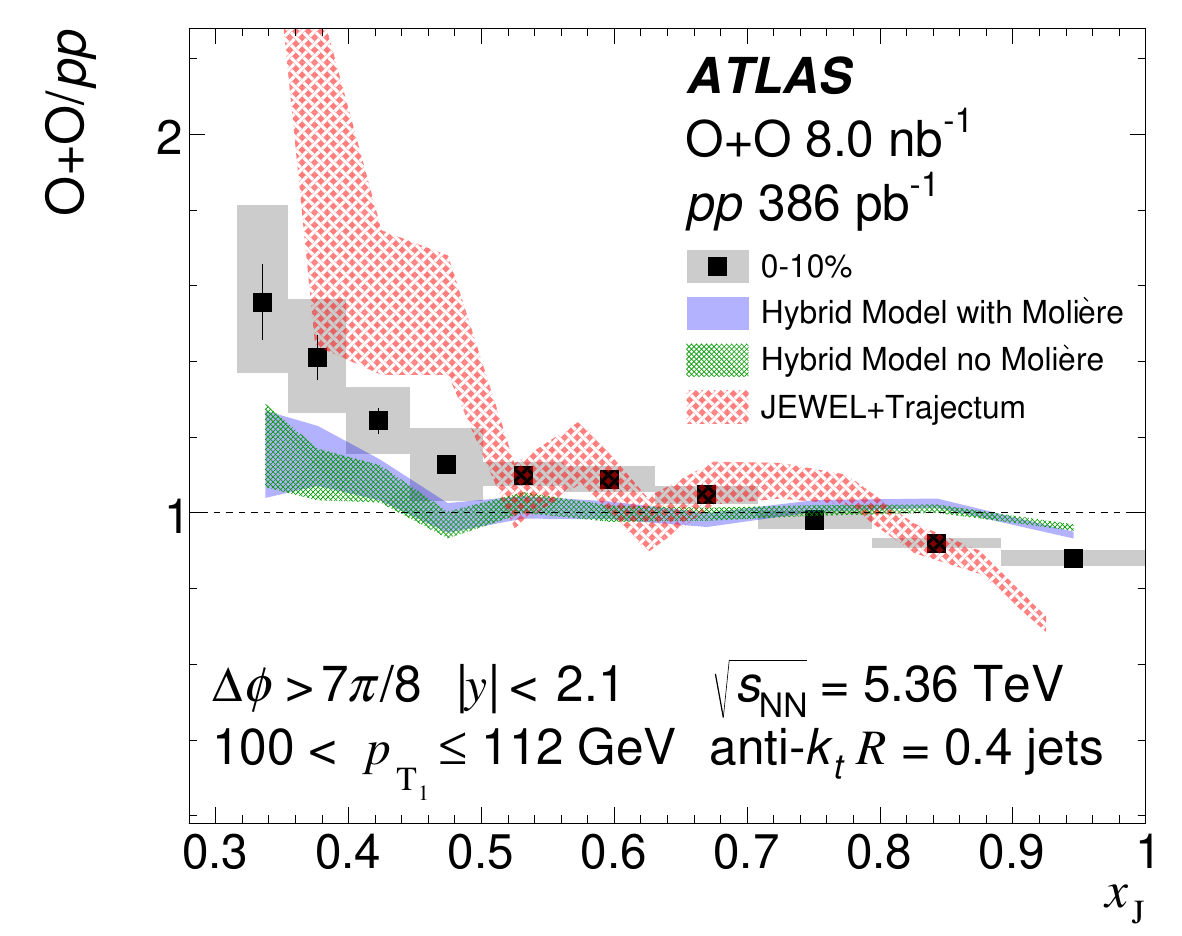}
(b)\\[-1ex]
\end{minipage}

\caption{
Ratios of unfolded \XJ\ distributions in \OxO\ to \pxp\
collisions for (a) $79< \PTONE \leq 89$~\GeV\ and
(b) $100< \PTONE \leq 112$~\GeV.
Ratios using central \OO\ collisions are represented by the
black markers. Vertical error bars represent the statistical uncertainty, and the shaded areas represent the total uncertainty.
The results are compared with predictions from the
\textsc{Hybrid} model with and without Moli\`ere scatterings~\cite{HybridModelOO,HybridModelOOMoreMoliere}
and the \textsc{Jewel+Trajectum} model~\cite{KolbePrediction}, with medium recoil disabled in \textsc{Jewel} in the presented calculation.
}
\label{fig:xjRatio_OOhybrid_suppEM}
\end{figure}

Figure~\ref{fig:xjRatio_OOhybrid_suppEM} compares the \XJ\ ratios in central \OO\ collisions to predictions from the \textsc{Hybrid}~\cite{HybridModelOO,HybridModelOOMoreMoliere} and \textsc{Jewel+Trajectum}~\cite{KolbePrediction} models. The \textsc{Hybrid} model combines perturbative QCD calculations with interactions between jet constituents and the QGP medium, while \textsc{Jewel+Trajectum} couples the \textsc{Jewel} jet-quenching framework to event-by-event hydrodynamic backgrounds provided by \textsc{Trajectum}, with medium recoil disabled in the presented calculation. For the \textsc{Hybrid} model, predictions are shown with and without Moli\`ere scatterings, whose relative contribution to jet energy loss is expected to increase in smaller collision systems due to their weaker path-length dependence. Both models predict a modification of the \XJ\ distribution in central \OO\ collisions, with the \textsc{Hybrid} (\textsc{Jewel+Trajectum}) model generally underestimating (overestimating) the magnitude of the effect observed in the data. The inclusion of Moli\`ere scatterings in the \textsc{Hybrid} model produces only small changes.

%

%
%
%

%
%
%
\FloatBarrier

\FloatBarrier
This Letter presented measurements of the dijet momentum balance, \XJ, in \OO\ and \NeNe\ collisions at $\sqrt{s_{\mathrm{NN}}}=5.36$~\TeV\ for five centrality intervals, compared with \pp\ collisions at the same energy.
The measurement is performed for leading jets between 63 and 251~\GeV,
with dijet pairs selected with $\Delta\phi > 7\pi/8$,
and both jets having $|y| < 2.1$.
The \XJ\ distributions are unfolded for detector effects, and the distributions
are normalized by the yield of dijets in each leading jet selection.
An increase in the relative contribution of lower \XJ\ dijets in 0--10\% central \OO\ and \NeNe\ collisions compared with both \pp\ and peripheral collisions for each light ion system is observed over the range of leading jet \pt\ reported in this measurement.
The significance of the modification is quantified by comparing the ratios of the \XJ\ distributions in \OO\ and \NeNe\ collisions to those in \pp\ collisions with the null hypothesis unity, corresponding to no modification. %
The modification is largest in the most central collisions and remains statistically significant in \OO\ (\NeNe) collisions up to the 20--40\% (10--20\%) centrality interval. In 0--10\% central \OO\ (\NeNe) collisions, significant deviations from the no-modification hypothesis are observed up to \PTONE\ 141 (126) GeV.

The observed modifications of the \XJ\ distributions are qualitatively similar to those previously measured in nucleus--nucleus collisions~\cite{HION-2020-09,HION-2018-28}, while occurring in a regime where projectile-configuration effects~\cite{Alvioli:2017wou,HION-2023-05} and event-activity selection biases~\cite{Park:2025mbt} are explicitly constrained.
These measurements provide new constraints on the onset and path-length dependence of jet quenching in small collision systems, extending the study of partonic energy loss toward the smallest quark--gluon plasma droplets accessible at the LHC.


%

%
%
%
%
%
\section*{Acknowledgements}
%

%

%
%

%
%

We thank CERN for the very successful operation of the LHC and its injectors, as well as the support staff at
CERN and at our institutions worldwide without whom ATLAS could not be operated efficiently.

The crucial computing support from all WLCG partners is acknowledged gratefully, in particular from CERN, the ATLAS Tier-1 facilities at TRIUMF/SFU (Canada), NDGF (Denmark, Norway, Sweden), CC-IN2P3 (France), KIT/GridKA (Germany), INFN-CNAF (Italy), NL-T1 (Netherlands), PIC (Spain), RAL (UK) and BNL (USA), the Tier-2 facilities worldwide and large non-WLCG resource providers. Major contributors of computing resources are listed in Ref.~\cite{ATL-SOFT-PUB-2026-001}.

We gratefully acknowledge the support of ANPCyT, Argentina; YerPhI, Armenia; ARC, Australia; BMWFW and FWF, Austria; ANAS, Azerbaijan; CNPq and FAPESP, Brazil; NSERC, NRC and CFI, Canada; CERN; ANID, Chile; CAS, MOST and NSFC, China; Minciencias, Colombia; MEYS CR, Czech Republic; DNRF and DNSRC, Denmark; IN2P3-CNRS and CEA-DRF/IRFU, France; SRNSFG, Georgia; BMFTR, HGF and MPG, Germany; GSRI, Greece; RGC and Hong Kong SAR, China; ICHEP and Academy of Sciences and Humanities, Israel; INFN, Italy; MEXT and JSPS, Japan; CNRST, Morocco; NWO, Netherlands; RCN, Norway; MNiSW, Poland; FCT, Portugal; MNE/IFA, Romania; MSTDI, Serbia; MSSR, Slovakia; ARIS and MVZI, Slovenia; DSI/NRF, South Africa; MICIU/AEI, Spain; SRC and Wallenberg Foundation, Sweden; SERI, SNSF and Cantons of Bern and Geneva, Switzerland; NSTC, Taipei; TENMAK, T\"urkiye; STFC/UKRI, United Kingdom; DOE and NSF, United States of America.

Individual groups and members have received support from BCKDF, CANARIE, CRC and DRAC, Canada; CERN-CZ, FORTE and PRIMUS, Czech Republic; COST, ERC, ERDF, Horizon 2020 and Marie Sk{\l}odowska-Curie Actions, European Union; Investissements d'Avenir Labex, Investissements d'Avenir Idex and ANR, France; DFG and AvH Foundation, Germany; Herakleitos, Thales and Aristeia programmes co-financed by EU-ESF and the Greek NSRF, Greece; BSF-NSF and MINERVA, Israel; NCN and NAWA, Poland; La Caixa Banking Foundation, CERCA and AGAUR programs from Generalitat de Catalunya and PROMETEO and GenT Programmes Generalitat Valenciana, Spain; G\"{o}ran Gustafssons Stiftelse, Sweden; The Royal Society and Leverhulme Trust, United Kingdom; Eric and Wendy Schmidt Fund for Strategic Innovation, United States of America.

In addition, individual members wish to acknowledge support from Chile: Agencia Nacional de Investigaci\'on y Desarrollo (ANID FONDECYT reg. 1230987, FONDECYT 1230812, FONDECYT 1240864, Fondecyt 3240661, Fondecyt Regular 1240721); China: Chinese Ministry of Science and Technology (MOST-2023YFA1605700, MOST-2023YFA1609300), National Natural Science Foundation of China (NSFC 12275265, NSFC-W2543005); Czech Republic: Czech Science Foundation (GACR - 24-11373S), Ministry of Education Youth and Sports (ERC-CZ-LL2327, FORTE CZ.02.01.01/00/22\_008/0004632); EU: H2020 European Research Council (ERC - 101002463); European Union: European Research Council (BARD No. 101116429, ERC - 101219398, ERC 101089007), European Regional Development Fund (HE COFUND GA No.101081355, ERDF), Marie Sklodowska-Curie Actions (GAP-101168829); France: Agence Nationale de la Recherche (ANR-21-CE31-0013, ANR-22-EDIR-0002, ANR-24-CE31-0504-01); Germany: Deutsche Forschungsgemeinschaft (DFG - 469666862); China: Research Grants Council (GRF); Italy: Istituto Nazionale di Fisica Nucleare (LHC-MIUR - 28003/2025), Ministero dell'Università e della Ricerca (NextGenEU I53D23000820006 M4C2.1.1, SOE2024\_0000023); Japan: Japan Society for the Promotion of Science (JSPS KAKENHI  JP25H0063, JSPS KAKENHI JP22H04944, JSPS KAKENHI JP24K23939, JSPS KAKENHI JP24KK0251, JSPS KAKENHI JP25H00650, JSPS KAKENHI JP25H01291, JSPS KAKENHI JP25K01011, JSPS KAKENHI JP25K01023, JSPS KAKENHI JP25KK0047); Poland: Polish National Science Centre (NCN 2021/42/E/ST2/00350, NCN OPUS 2023/51/B/ST2/02507, NCN OPUS nr 2022/47/B/ST2/03059, NCN UMO-2019/34/E/ST2/00393, UMO-2022/47/O/ST2/00148, UMO-2023/49/B/ST2/04085, UMO-2023/51/B/ST2/00920, UMO-2024/53/N/ST2/00869); Spain: Agència de Gestió d'Ajuts Universitaris i de Recerca. (AGAUR - 2023 BP 00141), Ministry of Science and Innovation (RYC2019-028510-I, RYC2020-030254-I, RYC2021-031273-I, RYC2022-038164-I), Ministerio de Ciencia, Innovación y Universidades/Agencia Estatal de Investigaci\'on (EU NextGenerationEU (PRTR-C17.I1), PID2022-142604OB-C22); Sweden: Carl Trygger Foundation (Carl Trygger Foundation CTS 22:2312), Swedish Research Council (Swedish Research Council 2023-04654, VR 2021-03651, VR 2022-03845, VR 2022-04683, VR 2023-03403, VR 2024-05451, VR 2025-05940), Knut and Alice Wallenberg Foundation (KAW 2023.0366); United Kingdom: The Binks Trust, Royal Society (NIF-R1-231091); United States of America: U.S. Department of Energy (ECA DE-AC02-76SF00515), John Templeton Foundation (John Templeton Foundation 63206), Neubauer Family Foundation.

%
%


%
\clearpage
\part*{End Matter}

\subsection*{\textit{Appendix A: Comparison to peripheral \pbpb}}

\begin{figure}[ht]
\centering

\begin{minipage}{0.49\linewidth}
\centering
\includegraphics[width=\linewidth,
trim={0cm 0cm 0cm 0mm},clip]{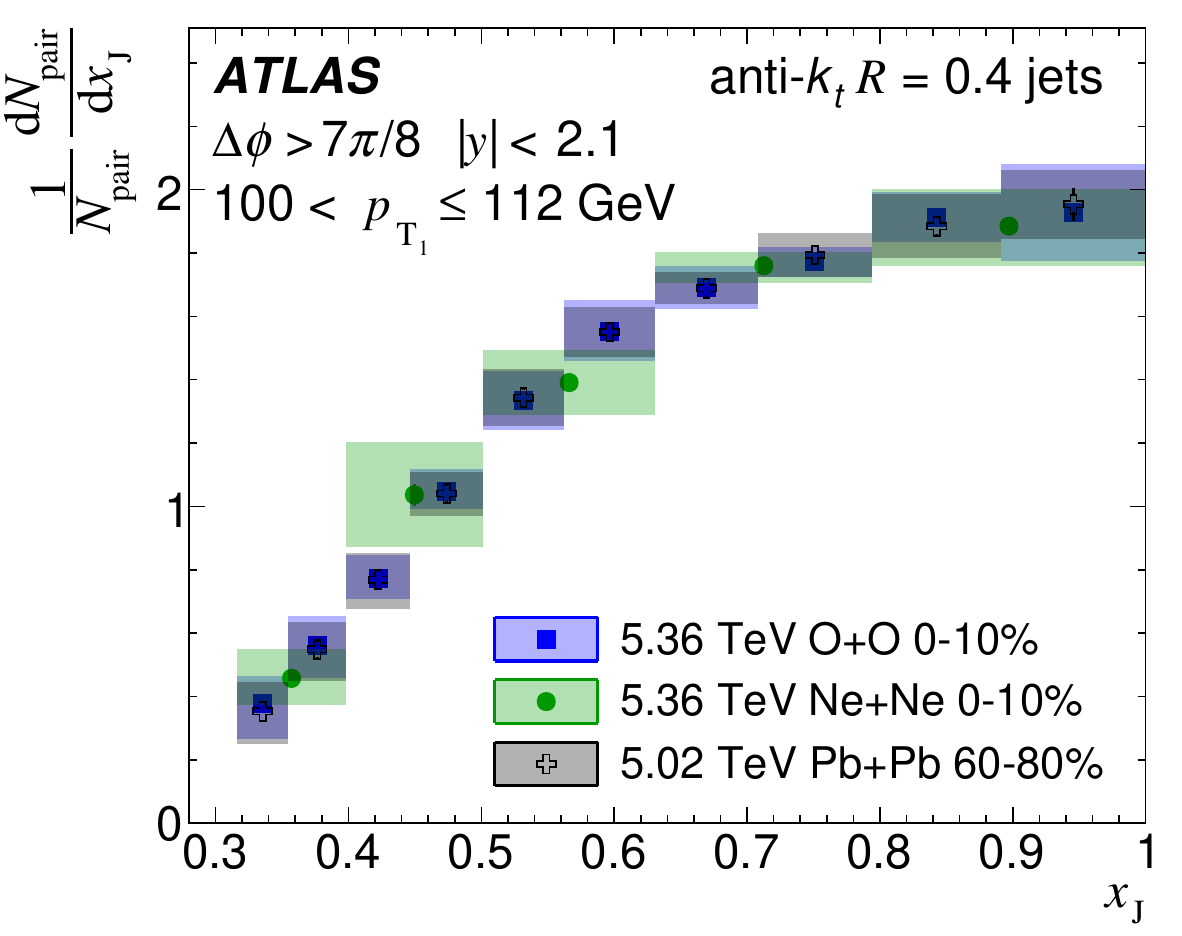}
(a)\\[-1ex]
\end{minipage}
\hfill
\begin{minipage}{0.49\linewidth}
\centering
\includegraphics[width=\linewidth,
trim={0cm 0cm 0cm 0mm},clip]{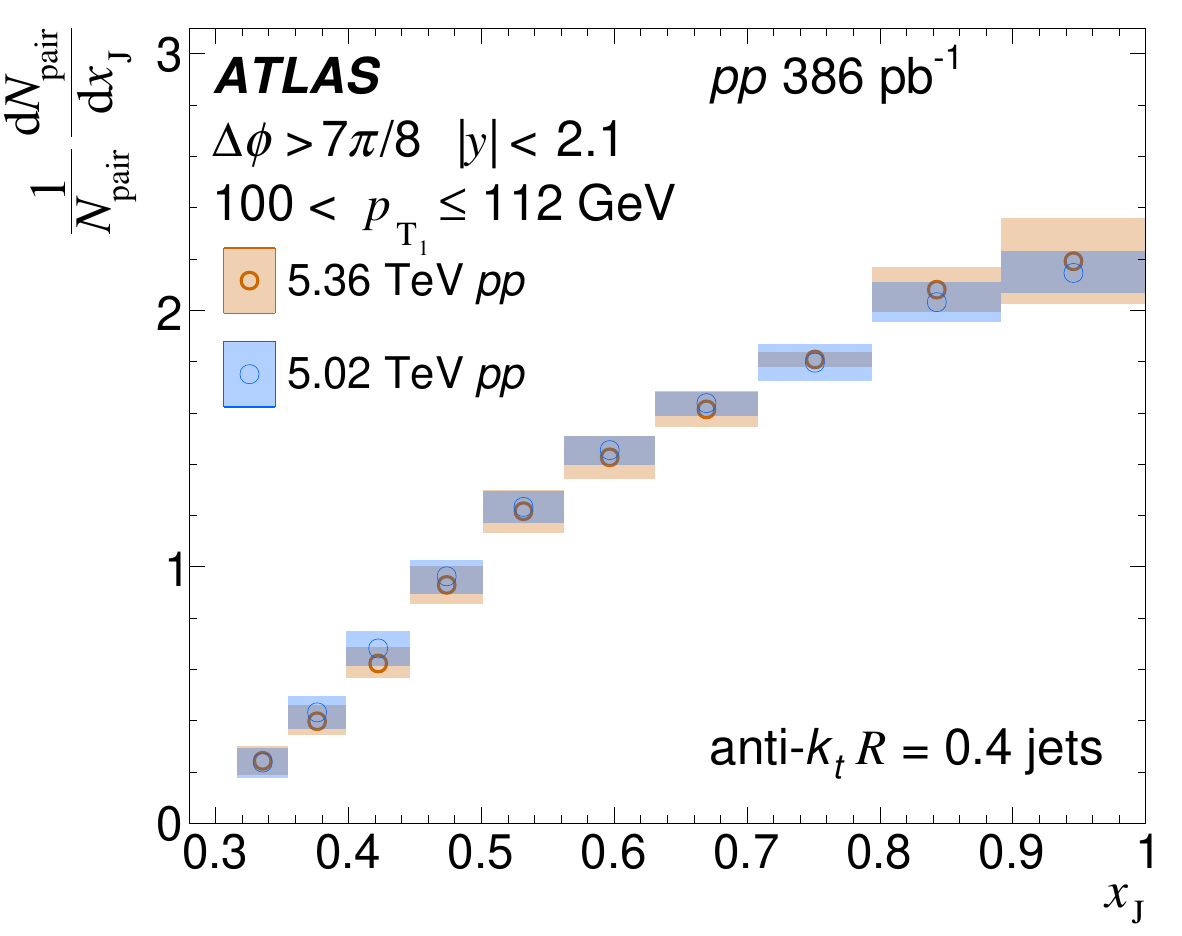}
(b)\\[-1ex]
\end{minipage}

\caption{
(a) Unfolded \XJ\ distributions in 0--10\% central \OxO,
\NeNe, and 60--80\% peripheral \pbpb\ \cite{HION-2019-02} collisions in a common bin
of leading-jet \PT. The light-ion centrality intervals correspond
to a similar level of event activity as the selected \pbpb\
centrality interval.
(b) Corresponding \pxp\ measurements at the relevant
center-of-mass energies.
}
\label{fig:xj_system_comp}
\end{figure}

The \XJ\ distributions measured in \OO\ and \NeNe\ collisions are compared with a previous measurement in \pbpb\ collisions in Figure~\ref{fig:xj_system_comp}(a).
To compare the modifications observed in \OO, \NeNe, and \pbpb\ collisions, the event activity measured by \fcalsum\ is roughly matched between the three systems~\cite{HION-2025-02,HION-2018-28}. The 0--10\% most-central \OO\ and  \NeNe\ collisions correspond to an event activity comparable to that of peripheral \pbpb\ collisions (60--80\%)~\cite{HION-2019-02}. At similar event activity, the distributions are found to be comparable between the three systems.
The \OO\ and \NeNe\ measurements were performed at slightly higher center-of-mass energy ($\SNN=5.36$~\TeV) than the \pbpb\ measurement ($\SNN=5.02$~\TeV). A comparison between the corresponding \pp\ measurements at the two energies was carried out, demonstrating no significant energy dependence of the \XJ\ observable within uncertainties, as shown in Figure~\ref{fig:xj_system_comp}(b).

\subsection*{\textit{Appendix B: Significance of the modification of the \XJ\ distributions}}

The significance of the modification of the \XJ\ distributions in \OO\ and \NeNe\ collisions relative to \pp\ collisions is quantified using a $\chi^2$ test of the ratio to the null hypothesis of no modification (unity). The test statistic is defined as
\begin{equation}
\chi^2 = \left(\mathbf{R}-\mathbf{I}\right)^{\mathrm{T}}
\mathbf{V}^{-1}
\left(\mathbf{R}-\mathbf{I}\right),
\end{equation}
where $\mathbf{R}$ is the ratio of the \XJ\ distributions, $\mathbf{I}$ is the unity vector, and $\mathbf{V}$ is the total covariance matrix including statistical and systematic bin-to-bin correlations after self-normalization. For asymmetric uncertainties, the covariance matrix corresponding to the uncertainty variation that yields the smallest deviation from unity is used.

The p-value obtained from the $\chi^2$ test is converted to a Z-score significance. The significance values for all \OO\ and \NeNe\ centrality intervals are reported in Tables~\ref{tab:OOsignificance} and~\ref{tab:NeNesignificance}. In both systems, significances above $5\sigma$ are observed in central collisions for leading-jet \PT\ below 100~\GeV. Significant modifications persist up to the 20--40\% centrality interval in \OxO\ and the 10--20\% centrality interval in \NeNe, while the significance decreases toward more peripheral collisions and higher leading-jet \PT.

\begin{table}[!ht]
\centering
\caption{$\chi^2$ test of the ratio of \XJ\ distributions in \OxO\ to \pxp\ relative to the null hypothesis of no modification. The corresponding significance values are reported for each leading-jet \PT\ and centrality interval. Significances are obtained by converting the $\chi^2$ $p$-value into an equivalent one-sided Gaussian significance. Entries reported as 0.0 correspond to p-values larger than 0.5, for which the equivalent one-sided Gaussian significance is not defined on the signal-like tail and is therefore set to zero.}
\vspace{2mm}
\begin{tabular}{ | c | c | c | c | c | c | }
\hline
\multicolumn{6}{| c |}{All Centralities \OxO\ Significance}\\
\hline
\rule{0pt}{2.5ex}
\hspace{2mm} \textbf{\PTONE} [GeV] & \hspace{2mm} 0--10\% & \hspace{2mm} 10--20\% & \hspace{2mm} 20--40\% & \hspace{2mm} 40--60\% & \hspace{2mm} 60--80\% \\[0.5ex]
\hline
\hline[63, 71]   & 9.8   & 2.5   & 6.3   & 1.4   & 0.0              \\\hline[71, 79]   & 6.4   & 6.8   & 4.0   & 0.0   & 0.0                   \\\hline[79, 89]   & 10.2  & 7.4   & 5.8   & 0.6   & 0.0                      \\\hline[89, 100]  & 9.3   & 5.1   & 5.3   & 0.0   & 0.0                    \\\hline[100, 112] & 7.3   & 7.4   & 1.0   & 0.0   & 0.0                 \\\hline[112, 126] & 5.5   & 4.5   & 2.1   & 0.4   & 0.0                     \\\hline[126, 141] & 5.2   & 1.8   & 0.1   & 0.0   & 0.0             \\\hline[141, 251] & 4.7   & 3.2   & 0.0   & 0.0   & 0.0                     \\\hline
\end{tabular}
\label{tab:OOsignificance}
\end{table}

\begin{table}[!ht]
\centering
\caption{$\chi^2$ test of the ratio of \XJ\ distributions in \NeNe\ to \pxp\ relative to the null hypothesis of no modification. The corresponding significance values are reported for each leading-jet \PT\ and centrality interval. Significances are obtained by converting the $\chi^2$ $p$-value into an equivalent one-sided Gaussian significance. Entries reported as 0.0 correspond to p-values larger than 0.5, for which the equivalent one-sided Gaussian significance is not defined on the signal-like tail and is therefore set to zero.}
\vspace{2mm}
\begin{tabular}{ | c | c | c | c | c | c | }
\hline
\multicolumn{6}{| c |}{All Centralities \NeNe\ Significance}\\
\hline
\rule{0pt}{2.5ex}
\hspace{2mm} \textbf{\PTONE} [GeV] & \hspace{2mm} 0--10\% & \hspace{2mm} 10--20\% & \hspace{2mm} 20--40\% & \hspace{2mm} 40--60\% & \hspace{2mm} 60--80\% \\[0.5ex]
\hline
\hline[63, 71]     & 11.5   & 6.6   & 4.3   & 1.3   & 0.4                      \\\hline[71, 79]     & 9.2   & 4.8   & 2.8   & 1.1   & 0.3                         \\\hline[79, 89]     & 8.9   & 5.3   & 4.9   & 0.3   & 0.0                        \\\hline[89, 100]    & 6.2   & 6.0   & 4.3   & 0.7   & 0.0                          \\\hline[100, 112]   & 6.3   & 3.6   & 3.3   & 0.0   & 0.0                         \\\hline[112, 126]   & 6.7   & 1.5   & 2.3   & 0.0   & 0.0                      \\\hline[126, 141]   & 4.0   & 1.6   & 1.8   & 0.0   & 0.0                   \\\hline[141, 251]   & 4.3   & 3.2   & 0.9   & 0.0  & 0.0                      \\\hline
\end{tabular}
\label{tab:NeNesignificance}
\end{table}


%

%
%

%
%
\printbibliography
\clearpage
\input{atlas_authlist}
%

%
%

%
%

%
%
%
%

\end{document}

%% file: atlas_authlist.tex
 
\begin{flushleft}
\hypersetup{urlcolor=black}
{\Large The ATLAS Collaboration}

\bigskip

\AtlasOrcid[0000-0002-6665-4934]{G.~Aad}$^\textrm{\scriptsize 102}$,
\AtlasOrcid[0000-0001-7616-1554]{E.~Aakvaag}$^\textrm{\scriptsize 17}$,
\AtlasOrcid[0000-0002-5888-2734]{B.~Abbott}$^\textrm{\scriptsize 121}$,
\AtlasOrcid[0000-0002-0287-5869]{S.~Abdelhameed}$^\textrm{\scriptsize 83b}$,
\AtlasOrcid[0000-0002-1002-1652]{K.~Abeling}$^\textrm{\scriptsize 54}$,
\AtlasOrcid[0000-0001-5763-2760]{N.J.~Abicht}$^\textrm{\scriptsize 48}$,
\AtlasOrcid[0000-0002-8496-9294]{S.H.~Abidi}$^\textrm{\scriptsize 30}$,
\AtlasOrcid[0009-0003-6578-220X]{M.~Aboelela}$^\textrm{\scriptsize 44}$,
\AtlasOrcid[0000-0002-9987-2292]{A.~Aboulhorma}$^\textrm{\scriptsize 36e}$,
\AtlasOrcid[0000-0001-5329-6640]{H.~Abramowicz}$^\textrm{\scriptsize 154}$
% ...
\bigskip
\\

% ...
$^{17}$Department for Physics and Technology, University of Bergen, Bergen; Norway.\\
% ...
$^{30}$Physics Department, Brookhaven National Laboratory, Upton NY; United States of America.\\
% ...
$^{36}$$^{(a)}$Facult\'e des Sciences Ain Chock, Universit\'e Hassan II de Casablanca;$^{(b)}$Facult\'{e} des Sciences, Universit\'{e} Ibn-Tofail, K\'{e}nitra;$^{(c)}$Facult\'e des Sciences Semlalia, Universit\'e Cadi Ayyad, LPHEA-Marrakech;$^{(d)}$LPMR, Facult\'e des Sciences, Universit\'e Mohamed Premier, Oujda;$^{(e)}$Facult\'e des sciences, Universit\'e Mohammed V, Rabat;$^{(f)}$Institute of Applied Physics, Mohammed VI Polytechnic University, Ben Guerir; Morocco.\\
% ...
$^{44}$Physics Department, Southern Methodist University, Dallas TX; United States of America.\\
% ...
$^{48}$Fakult\"{a}t Physik, Technische Universit{\"a}t Dortmund, Dortmund; Germany.\\
% ...
$^{54}$II. Physikalisches Institut, Georg-August-Universit\"{a}t G\"ottingen, G\"ottingen; Germany.\\
% ...
$^{83}$$^{(a)}$Khalifa University of Science and Technology, Abu Dhabi;$^{(b)}$New York University Abu Dhabi, Abu Dhabi;$^{(c)}$United Arab Emirates University, Al Ain;$^{(d)}$University of Sharjah, Sharjah; United Arab Emirates.\\
% ...
$^{102}$CPPM, Aix-Marseille Universit\'e, CNRS/IN2P3, Marseille; France.\\
% ...
$^{121}$Homer L. Dodge Department of Physics and Astronomy, University of Oklahoma, Norman OK; United States of America.\\
% ...
$^{154}$Raymond and Beverly Sackler School of Physics and Astronomy, Tel Aviv University, Tel Aviv; Israel.\\
% ...
$^{a}$ Also at Affiliated with an institute formerly covered by a cooperation agreement with CERN.\\
$^{b}$ Also at An-Najah National University, Nablus; Palestine.\\
$^{c}$ Also at Borough of Manhattan Community College, City University of New York, New York NY; United States of America.\\
$^{d}$ Also at Center for Interdisciplinary Research and Innovation (CIRI-AUTH), Thessaloniki; Greece.\\
$^{e}$ Also at Centre of Physics of the Universities of Minho and Porto (CF-UM-UP); Portugal.\\
$^{f}$ Also at CERN, Geneva; Switzerland.\\
% ...

\end{flushleft}
